\documentclass[fleqn,12pt,twoside]{article}
\usepackage[headings]{espcrc1}

\usepackage{graphicx}
\usepackage[figuresright]{rotating}

\newcommand{\AmS}{{\protect\the\textfont2
  A\kern-.1667em\lower.5ex\hbox{M}\kern-.125emS}}

\title{Evolution of the concept of  Quark Matter: 
the Ianus face of the heavy ion collisions}
\author{J. Zim\'anyi\address[RMKI]{KFKI Research Institute for Nuclear and 
          Particle Physics, \\
          P.O. Box 49, H-1525 Budapest, Hungary}
        \thanks{The author thanks for the financial support of the OTKA 
         grant T 49466. The author also thanks to the STAR collaboration for the 
	 permission of reproducing their result in Fig.\ref{Fig.7}.}
        }

\runtitle{Evolution of the concept of  Quark Matter}
\runauthor{J. Zim\'anyi}

\begin{document}

\maketitle

\begin{abstract}
Since the beginning of the efforts to produce and understand 
quark matter large changes developed in the ideas of description
of this matter. In the present paper we summarize some aspects
of this development.
\end{abstract}


\section{Evolution of ideas on quark matter}

Since the beginning of the quest of quark matter deep changes have developed 
in our concept in understanding its properties. 
In Table \ref{Table.1} the ideas are collected, which dominated 
at the beginning of quark matter research. The present status of these
ideas are also displayed. 
Although theoretical models supported the early concept, but the
experimental data forced to change these speculations step by step.
Thus the picture of a static, weakly interacting gaseous quark-gluon plasma
was substituted by the more dynamic, strongly interacting fluid-like
quark-gluon matter, where the ingredients gained an effective mass
during the intense microscopical interactions.

\vspace*{-0.3truecm}
\begin{table}[htb]
\newcommand{\m}{\hphantom{$-$}}
\newcommand{\cc}[1]{\multicolumn{1}{c}{#1}}
\renewcommand{\tabcolsep}{2pc} 
\renewcommand{\arraystretch}{1.2} 
\begin{tabular}{@{}ll}
\hline
{\bf Early times}   & 
  {\bf Recent status} \\
   \hline\hline 
{very high quark density;}  &
  {not too high quark density;} \\ 
{vanishing coupling constant;} & 
  {large effective coupling constant;} \\ 
{zero mass for quarks and gluons;}   &   
  {finite effective mass for quarks and gluons;} \\  
{weakly interacting constituents; } & 
{strongly interacting constituents; }  \\
\hline
  {static, homogeneous  gas;} &
  {dynamic, liquid-like system} \\ 
 {very high temperature}; &
  {temperature near to critical}; \\
 {vanishing sound velocity}; &
  {finite sound velocity}; \\
 {long lifetime}; &
  {short lifetime}; \\
\hline
\end{tabular}\\[2pt]
\caption{Concept of Quark Matter in the early times \cite{qgp} and
recently, considering the microstructure (upper part of the table) 
and the bulk properties (lower part of the table). }
\label{Table.1}
\end{table}  

\newpage

\section{Time evolution of a heavy ion reaction}

As soon as the static picture of quark gluon plasma was substituted by a
dynamic scenario with short lifetime, the different pre-QGP and
post-QGP stages became as important as the QGP state itself.
As the dominance of a static plasma state has been lost, the dynamical
steps and the intermediate phases ('matters') became the target of the
theoretical investigations. Table 2 displays the emerged scenarios appearing 
in heavy ion collisions.
\begin{table}[htb]
\newcommand{\m}{\hphantom{$-$}}
\newcommand{\cc}[1]{\multicolumn{1}{c}{#1}}
\renewcommand{\tabcolsep}{2pc} 
\renewcommand{\arraystretch}{1.2} 
\begin{tabular}{@{}ll}
\hline
\hline  
 {\bf type of event}   & {\bf microscopic dynamics }   \\
   \hline\hline 
{\bf color glass condensate (?)} & gluon saturation~\cite{GyL}   \\
{\bf evolution of plasma}     & expands and cools, \\
                              & undergoes through 
   a set of phases~\cite{GLV}   \\
{\bf prehadronization stage}   & quarks and antiquarks gather effective 
  mass~\cite{LH}  \\
{\bf hadronization}  & quarks and antiquarks coalesce into 
hadrons~\cite{ALC}  \\
\hline
\end{tabular}\\[2pt]
\caption{Time evolution of a heavy ion reaction. }
\label{Table.2}
\end{table}  

\vspace{-5mm}
{\it In a heavy ion collision not a single sort, 
well defined type of matter is formed,
but a set of different types of matter is created 
in the consecutive stages of the reaction. }

\section{Evolution of ideas on hadronization }

In case of a quasi-static quark gluon plasma with long lifetime, there 
should be enough time for slow hadronization, which is driven by
energy and momentum transformation at the macroscopical level.
This leads to the application of thermodynamics. 15 years ago
we were very much confident to expect a 
{thermodynamical first order phase transition}.
As the time scale of the reaction became shorter, the  idea of 
{mixed phase transition} appeared, handled with thermodynamical methods.
However, as it turned out that the evolution of a heavy ion reaction is much 
faster, the hadronization ideas too had to be  adjusted to this 
faster scenario.
Finally, so fast hadronization processes must have been considered, that
the macroscopical description was substituted by microscopical one
and the slow thermodynamical transformation has been substituted
by the fast process of coalescence of constituent quarks.

\section{The coalescence models}

The concept of quark coalescence has been evolved through different steps.
At first the idea of 
{quark number conservation} appeared and became 
a very useful tool~\cite{ALC}. Then the 
{linear vs. nonlinear coalescence} descriptions were 
investigated~\cite{BZBCL}.
Later on the  description was improved and the direct
{momentum conservation} at the microscopical level was 
considered in the coalescence processes~\cite{HGF1,HGF2,HGF3}. 
Finally {mass conservation} has also been studied to produce light hadrons from
heavy constituent quarks~\cite{ALCOR05}.

\subsection{ Coalescence hadronization with   
  constituent  quark number conservation }

For the description of hadrons the constituent quark model was quite
successful. In this approach the quarks are dressed, with  mass of
approximatively $ 300$ MeV. The question arises, that 
 at what stage of
hadronization is  this effective mass created? In the
 ALCOR model~\cite{ALC}  it is assumed that
this effective mass is created in the very last stage, 
in the prehadronization stage of the evolution
of fireball.

{\it The main assumptions of the ALCOR model:}
\begin{description}
\vspace*{-0.3truecm}
\item[A)]{ At the beginning of  hadronization the quarks  
are dressed constituent quarks.}\vspace*{-0.3truecm}
\item[B)]{ These quarks coalesce to form the hadrons. }
\vspace*{-0.3truecm}
\item[C)]{ The number of different quarks and antiquarks
is conserved during  hadronization. }
\vspace*{-0.3truecm}
\item[D)]{ The effective mass of gluons
is much higher than that of quarks near the critical temperature.
Thus the gluon degree of freedom is neglected in this
late period of the heavy ion reaction.}
\vspace*{-0.3truecm}
\item[E)]{
The number of a given type of  hadrons is 
proportional to the product of the numbers of
different quarks from which the hadron consists:
\vspace*{-0.2truecm}
\begin{equation}
  N_{B,(ijk)} = C_{B,(ijk)} \, \cdot \,
(b_i \cdot N_i)\cdot(b_j \cdot N_j)\cdot(b_k \cdot N_k) 
\end{equation}
\vspace*{-0.2truecm}
where $N_{B,(ijk)} $ is the number of produced baryons from 
quark $ i,j,k $, and the
equation for quark number conservation in the hadronization 
     ${   N^{Hadronmatter}_{i}} = { N^{QM}_i = N_i }  $   
determines the $ b_i $ normalization coefficients.}
\end{description}

 With these assumptions 
 one gets the relative numbers 
of particles in good agreement with experimental data, see
Table \ref{Table.3}.

\begin{table}[htb]
\newcommand{\m}{\hphantom{$-$}}
\newcommand{\cc}[1]{\multicolumn{1}{c}{#1}}
\renewcommand{\tabcolsep}{2pc} 
\renewcommand{\arraystretch}{1.2} 
\begin{tabular}{@{}llll}
\hline
&ALCOR model & STAR data & Ref.  \\ \hline\hline
$h^- $        
& 280        & $280 \pm 20$      &\cite{STAR_h}    \\
${K^-}/{\pi^-} $ 
& 0.159      & $0.161 \pm 0.002$ &\cite{STAR_kpi} \\
${K^+}/{K^-} $ 
& 1.091      & $1.092 \pm 0.023$ &\cite{STAR_sas}\\ 
 ${\overline p}^-/{p}^+$
& 0.66       & $0.71 \pm 0.05$   &\cite{STAR_pap}  \\ 
 ${\overline \Lambda}/{\Lambda}$
& 0.72       & $0.71 \pm 0.01$   &\cite{STAR_sas}  \\ 
 ${\overline \Xi}^+/{\Xi}^-$
& 0.80       & $0.83 \pm 0.04$   &\cite{STAR_sas}  \\ 
 ${\Xi^-}/{h-} $
& 0.0091     & $0.0077 \pm 0.001$&\cite{STAR_xax}  \\
 ${\Phi}/{K^{*0}}$
& 0.42      & $0.49 \pm 0.12 $   &\cite{STAR_kstar}\\ \hline
\end{tabular}\\[2pt]
\caption{
Hadron production in Au+Au collision at $\sqrt{s}=130$ AGeV
from the ALCOR model and experimental data from the STAR
Collaboration~\cite{STAR_h,STAR_kpi,STAR_sas,STAR_pap,STAR_xax,STAR_kstar}.
}
\label{Table.3}
\end{table}

\subsection{ The simple quark counting }

Realizing~\cite{BZBCL} that the coalescence probabilities for a 
hadron and its antiparticle are the same:
                 $  C_{h} = C_{\overline{h}} $ ,
very transparent relations can be obtained 
for the ratios of the different multiplicities:

\begin{equation}
{\bf \frac{\overline \Lambda}{\Lambda} = D \cdot  \frac{\overline p}{p} } 
  \hspace{1.5cm} 
{\bf \frac{\overline \Xi}{\Xi} = D \cdot  \frac{\overline \Lambda}{\Lambda} } 
\hspace{1.5cm}
{\bf \frac{\overline \Omega}{\Omega} = D \cdot  \frac{\overline \Xi}{\Xi} }  
\hspace{1.5cm}
{\bf D = \frac{K}{\overline K} }
\label{simp2}
\end{equation}

\begin{figure}[h]  
{\includegraphics[width=80mm,height=75mm]{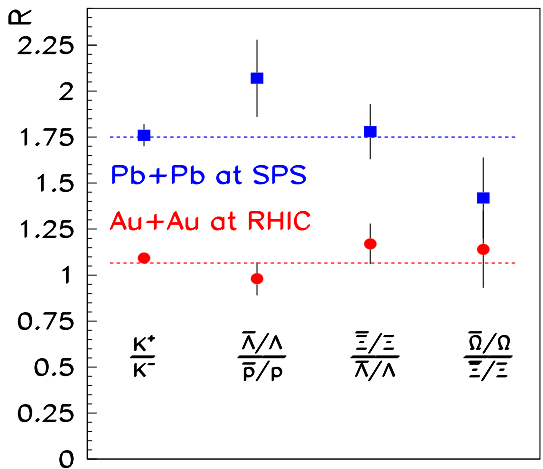}}
\caption[]{ 
Graphical display of the particle ratios at mid-rapidity
in {Au+Au collision at RHIC}~\cite{STAR_sas} and
in {Pb+Pb collisions at SPS}~\cite{NA49a,NA49b,NA49c,NA49d}. 
}
\label{Figure.1}
\end{figure}

These predictions agree again with the experimental data,
see Table \ref{Table.4} and  Figure \ref{Figure.1}.

\begin{table}[htb]
\newcommand{\m}{\hphantom{$-$}}
\newcommand{\cc}[1]{\multicolumn{1}{c}{#1}}
\renewcommand{\tabcolsep}{2pc} 
\renewcommand{\arraystretch}{1.4} 
\begin{tabular}{@{}lll}
\hline
 Ratios &  STAR & SPS \\ \hline\hline  
 $K^+/K^- $ 
         & { $1.092 \pm 0.023 $}  &{ $1.76 \pm 0.06 $}  \\
 $ {\{{\overline \Lambda}/{\Lambda}\}}/{\{{\overline p}/p\}} $  
     & $0.98 \pm 0.09$              &$ 2.07 \pm 0.21 $  \\ 
 $ {\{{\overline \Xi}/{\Xi}\}}/{\{{\overline \Lambda}/\Lambda\}} $
      & $1.17 \pm 0.11$              &$ 1.78 \pm 0.15$   \\
 $ {\{{\overline \Omega}/{\Omega}\}}/{\{{\overline \Xi}/\Xi\}} $
      & $1.14 \pm 0.21$              &$ 1.42 \pm 0.22$   \\  
\hline
\end{tabular}\\[2pt]
\caption{
Compilation of experimental particle ratios obtained at 
RHIC energy $\sqrt{s}=130$ AGeV~\cite{STAR_sas} 
and SPS energy $ 17.3 $ AGeV~\cite{NA49a,NA49b,NA49c,NA49d}.
}
\label{Table.4}
\end{table} 

\vspace*{-1.0truecm}

\subsection{Charge fluctuation in a quark-antiquark system }

The charge fluctuation~\cite{Bcf} is characterized by the ratio 
\begin{equation}
    D = 4 \frac{< \delta Q^2 >}{< N_{ch} >}  \ ,
\end{equation}
\noindent
where $ < \delta Q^2 > $ is the average of the charge fluctuation
and $ < N_{ch} > $ is the average value of the charge multiplicity.
The value of $D$
is  $ D \approx 3 $ for hadron gas in equilibrium,
and  $ D \approx 1 $ for the quark gluon plasma, 
The measured $ D $ value is near to 3. 
However, in the constituent quark coalescence scenario (ALCOR) one can
also expect a $ D \approx 3 $ value.

\subsection{ Coalescence 
with constituent quark momentum conservation} 

The idea of quark coalescence can be applied at relatively large
momenta and the hadron spectra at high-$p_T$ can be investigated
successfully~\cite{HGF1,HGF2,HGF3}. In this kinematic region,
especially at $p_T > 2$ GeV,
the mass of the light and strange quarks can be neglected,
  $ p_{quark} >> m_{quark}$, and the calculations are simplified. 
Assuming that the most dominant kinematic window is overwhelming, then
\begin{equation}
  { p_a = p_b = P_h / 2 \equiv P/2 } \\
\end{equation}

 The expression for the meson emission spectrum~\cite{HGF3}:
\begin{equation}
  \label{eq:mesfm}
  E \frac{d N_M}{d^3 P} = 
  \int\limits_{\Sigma_{f}} d\sigma \, \frac{P\cdot u(r)}{(2\pi)^3} \, 
  f_a(r;\frac{P}{2}) \, f_b(r;\frac{P}{2}) \,.\nonumber\\
\end{equation}

A similar expression is valid for the baryon emission spectrum:
\begin{equation}
  \label{eq:mesfb}
  E \frac{d N_B}{d^3 P} =
  \int\limits_{\Sigma_{f}} d\sigma \, \frac{P\cdot u(r)}{(2\pi)^3} \,
  f_a(r;\frac{P}{3}) \, f_b(r;\frac{P}{3}) \,f_c(r;\frac{P}{3}) .\nonumber\\
\end{equation}

These two equations lead to an interesting consequence: 
since  the momentum distribution of constituent quarks is expected 
to behave as an exponentially decreasing function of transverse momentum,
there are more partons at $ P/3 $ than at $ P/2 $.
\begin{equation}
 { f_q(r;\frac{P}{3}) > f_q(r;\frac{P}{2}) } 
\end{equation}

{} From that follows, that we have more baryons than mesons
 in the momentum range where hadrons are produced by coalescence.
This explains the surprising momentum dependence of the observed proton to 
meson ratio~\cite{HGF1,HGF2,HGF3}. This description also implies the
conservation of quark and antiquark numbers in the hadronization process.

\subsection{The effective mass of quarks}

{\it Mass distribution of quarks}

Since the effective mass of quarks are determined by their
interaction with the neighborhood,
and the matter in the fireball is not  smooth and homogeneous,
the granulated structure of the fireball
is reflected in the effective mass distribution of the quarks.

\noindent
The consequences of a possible quark mass distribution 
\begin{equation}
\rho(m)={\it N} e^
{ - \frac{\mu}{T_c} \sqrt{\frac{\mu}{m}+\frac{m}{\mu}} } 
\end{equation}  
 with $\mu=0.25$ GeV and $ T_c = 0.26$ GeV,  
are discussed in the next paragraph \cite{ALCOR05}.

\newpage

\begin{figure}[ht]
\begin{minipage}[t]{70mm}  
\includegraphics[width=65mm,height=90mm]{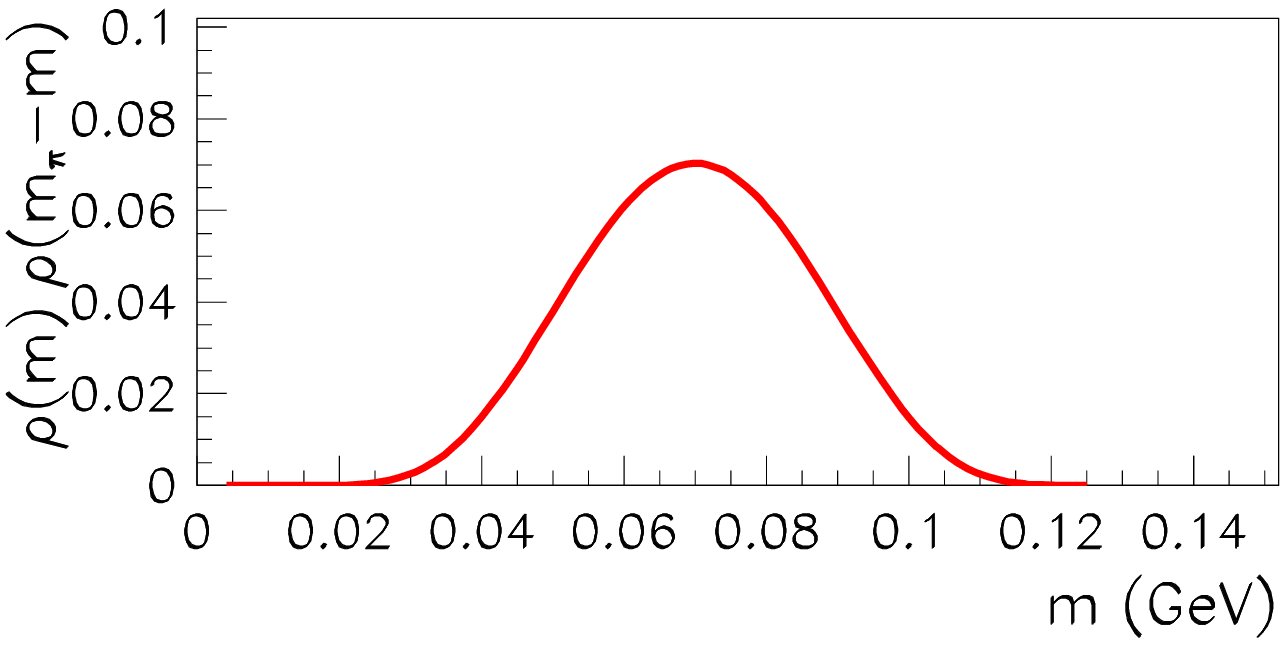}  
\vspace*{-45mm}
\caption{ The product of the two quark  mass distributions
         $\rho(m) \cdot \rho(m_{\pi} - m)$ in eq.(\ref{coalM}).
  The  maximum is at $ m=m_{\pi}/2 $.  }
\label{twooverlap1}
\end{minipage}
\hspace{\fill}
%
\begin{minipage}[t]{80mm} 
\vspace*{-88mm}
\includegraphics[width=45mm,angle=270]{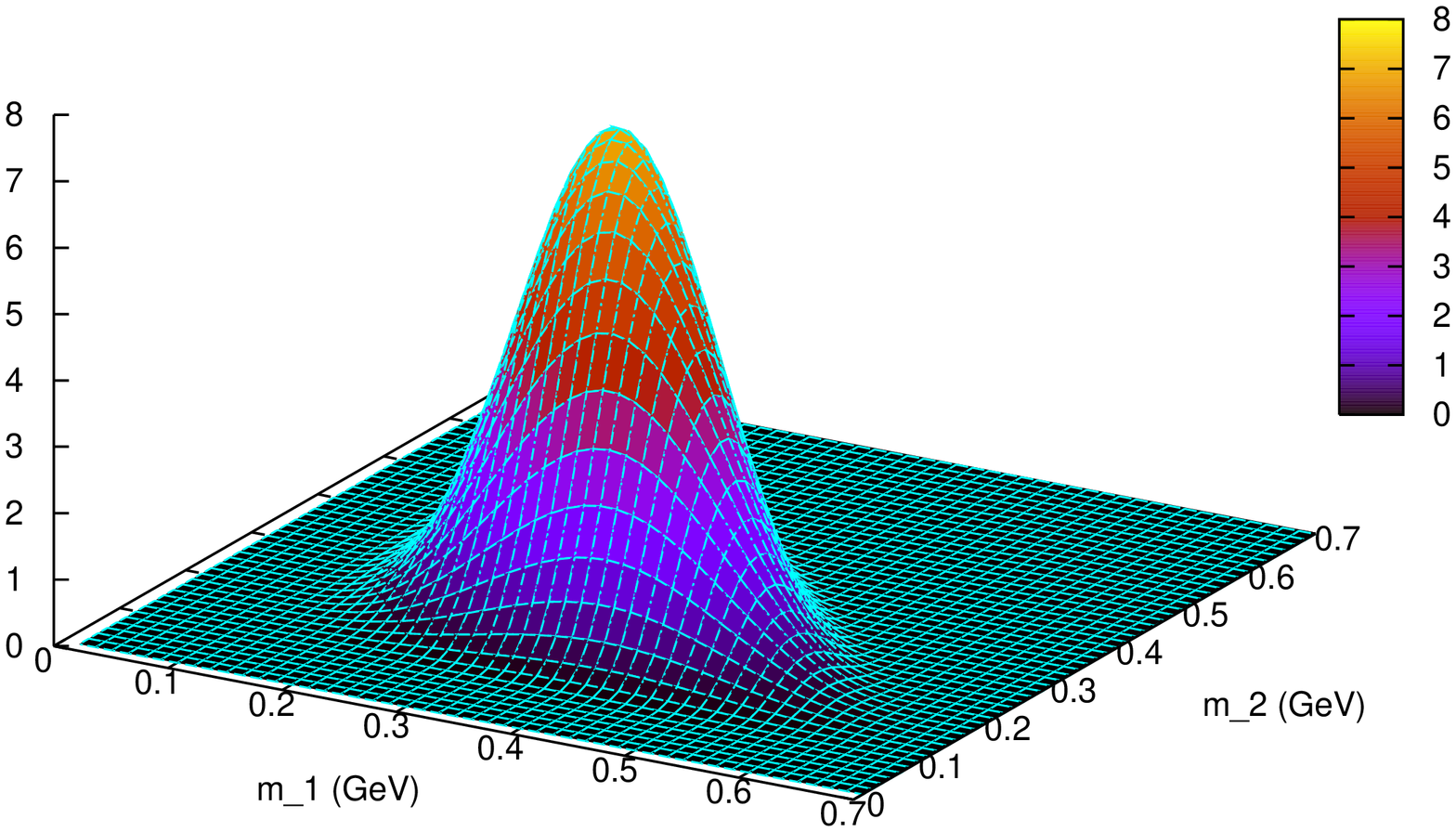}
\caption{Product of three mass distributions in eq.(\ref{coalB}). 
          The horizontal axes are  $m_1$ and $m_2$, the mass $m_3$
          is defined as  $m_3 = m_B-(m_1+m_2)$.  }
\label{threeoverlap1}
\end{minipage}
\end{figure}

\vspace*{-20mm}
\subsection{ The coalescence expression for distributed quark mass}

Using the covariant coalescence model of Dover {\it et al.},
 \cite{dover},
the spectra of hadrons formed from the coalescence of quark and antiquarks
can be written as
\begin{eqnarray}
E \cdot \frac{dN_h}{d^3{\bf p}} &=& 
 \frac{dN_h}{dy {\bf p}_T d{\bf p}_T d\phi_p} = 
 \frac{g_h}{(2\pi)^3}\int (p_h^{\mu} \cdot d\sigma_{h,\mu})
 \  { F_h( p_h; x_h)}  . 
\label{coal1}
\end{eqnarray}
Here $F_h(p_h; x_h)$ is an  eight dimensional distribution
(Wigner function) of the formed hadron, and
$d\sigma$ denotes an infinitesimal element of the space-like hypersurface
of hadron production. 

Assuming that meson production is homogeneous within the reaction volume,
the source function
for the $ a + b \rightarrow M $ meson production,
 $ F_M({\bf p}_M) $, is the following:
\begin{equation}
  F_M({\bf p}_M) = \int 
  d^3{\bf p}_a  d^3{\bf p}_b     
 \, f_a({\bf p}_a;0) \, f_b({\bf p}_b;0) \, 
C_M({\bf p}_a,{\bf p}_b,{\bf p}_M).
\label{coal4}
\end{equation}
The coalescence function $C_M({\bf p}_a,{\bf p}_b,{\bf p}_M)$  
is defined as
\begin{equation}
C_M({\bf p}_a,{\bf p}_b,{\bf p}_M) = 
\alpha_M \cdot  e^{-(({\bf p}_a - {\bf p}_M /2)/{P}_c)^2}  
         \cdot  e^{-(({\bf p}_b - {\bf p}_M /2)/{P}_c)^2}  
\label{coalfv}
\end{equation}
The parameters $\alpha_M$ and ${P}_c$ reflect properties of the hadronic
wave function in the momentum representation convoluted with the formation
matrix element.

Assume that  $P_c$ is so small, that partons with practically
 zero relative momentum  form a meson.
\begin{equation}
   {\bf p_a} = {\bf p_b} = {\bf p_M} / 2  \hspace{2.0cm} 
            m_a + m_b  = m_M  
\label{cons2}
\end{equation}
This  leads to the meson coalescence function
\begin{equation}
   C_M = \alpha_M \cdot {\bf \delta({\bf p}_a-{\bf p}_M/2)
 \cdot \delta({\bf p}_b-{\bf p}_M/2) } \nonumber
 \cdot \delta(m_a+m_b - m_M)  
\end{equation}
Thus we arrive at the following meson distribution function:  
\begin{eqnarray}
F_M({\bf p_t};0) &=&\alpha_M   
  \cdot \int_0^{m_M} dm_a  \cdot 
\int_0^{m_M} dm_b \cdot \delta(m_M -(m_a+m_b))  \nonumber \\
 &\cdot&  \rho(m_a) \cdot f_q({\bf p_t}/2;m_a) \ \ 
 \cdot \rho(m_b) \cdot f_b({\bf p_t}/2;m_b) \ .
\label{coalM}
\end{eqnarray}

\begin{figure}[ht]
\begin{minipage}[t]{75mm}
\includegraphics[width=75mm]{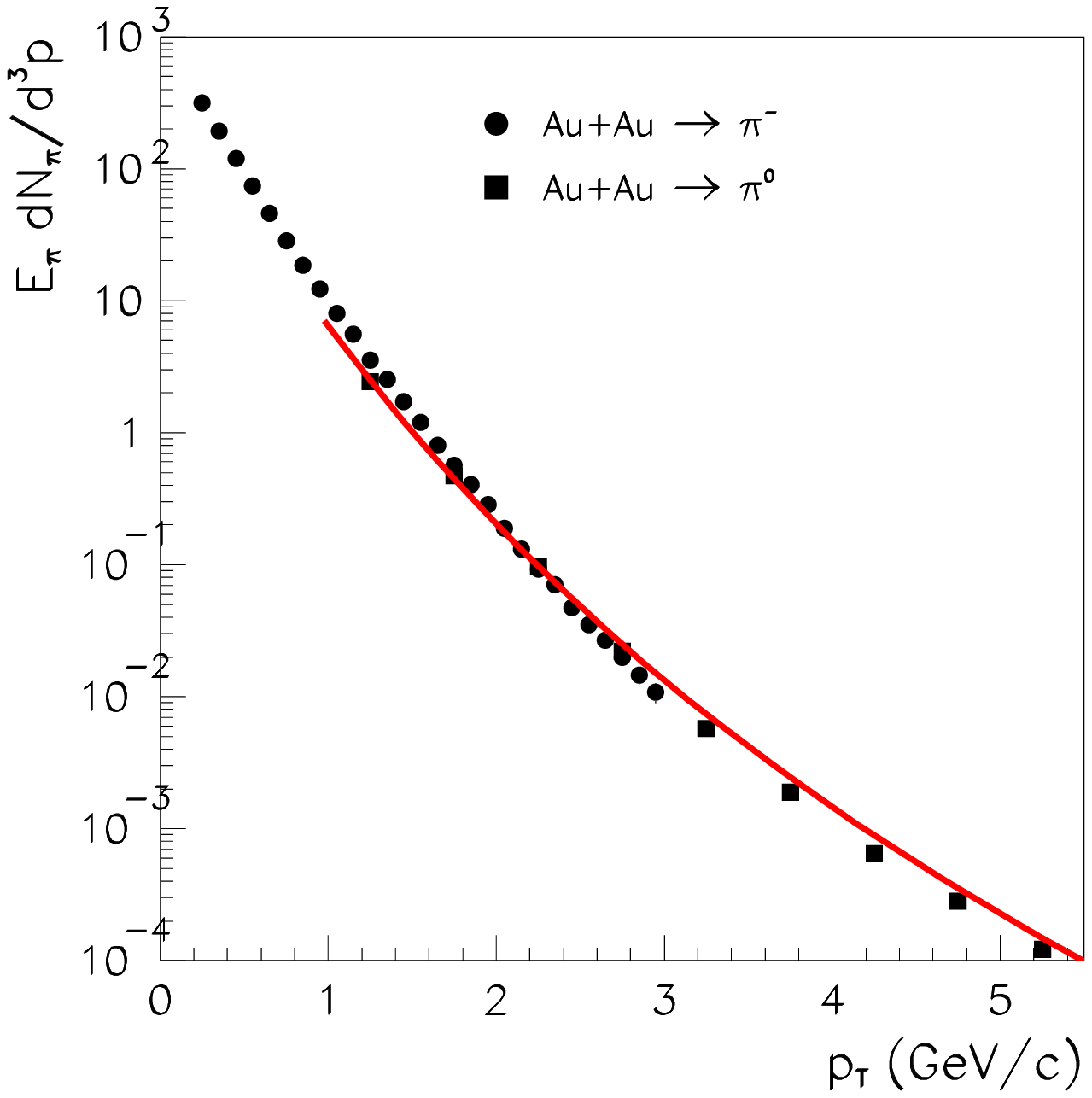}
\vspace{-13mm}
\caption{Measured \cite{PHENIX} and calculated $ {\pi}^{0} $ and $\pi^-$
         transvers momentum spectra in central Au+Au
          collisions at $\sqrt{s}=200$ AGeV. }
\label{spect_pion}
\end{minipage}
\hspace{\fill}
%
\begin{minipage}[t]{75mm} 
\includegraphics[width=75mm]{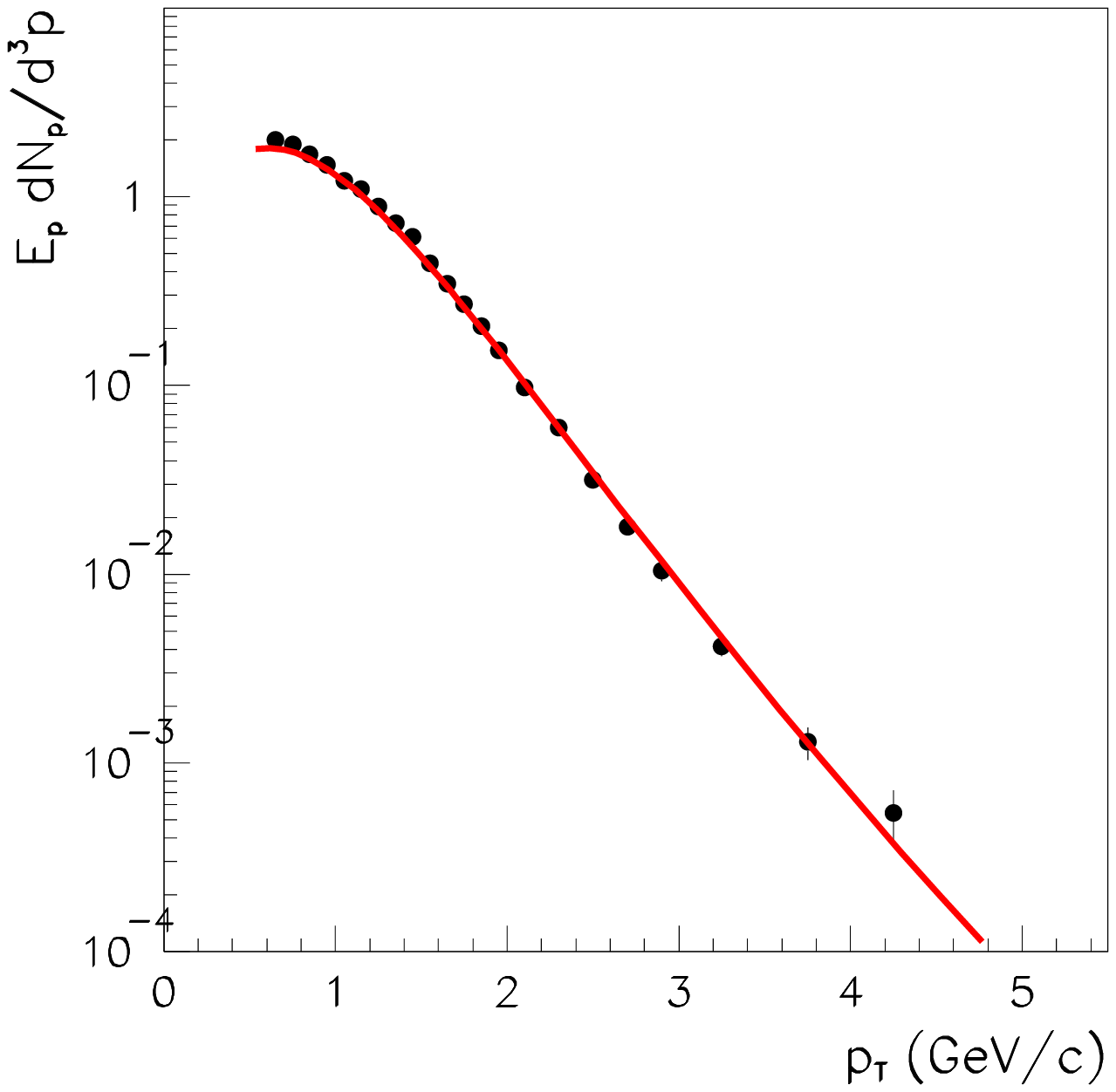}
\vspace{-13mm} 
\caption{Measured \cite{PHENIX} and calculated transverse
momentum spectrum for antiproton production
          in central Au+Au collisions at $\sqrt{s}=200$ AGeV.  }
\label{spect_aprot}
\end{minipage} 
\end{figure}

Similar argumentation leads to the three-fold coalescence expression. 
In the place of
 eq.(\ref{cons2}) we get:  
\begin{equation}
  {\bf  p}_1 = {\bf p}_2 = {\bf p}_3 = {\bf p}_B / 3 \hspace{10mm} 
     m_1 +m_2 +m_3 =  m_B 
\end{equation} 
and the baryon source function becomes
\begin{eqnarray}
  F_{B}({\bf p}_t)  
     &=& \alpha_B  
\int_0^{m_{pr}}\!\!\!dm_1 \int_0^{m_{pr}}\!\!\!dm_2 \int_0^{m_{pr}}\!\!\!dm_3 
\ \ \delta( m_B - (m_1+m_2+m_3) \ \cdot \  \nonumber \\
     && \cdot  \rho(m_1) \ f_{q}({\bf p}_t/3,m_1)  \cdot 
     \rho(m_2) \ f_{q}({\bf p}_t/3,m_2) \cdot  
     \rho(m_3) \ f_{q}({\bf p}_t/3,m_3)    \ \ .
\label{coalB}
\end{eqnarray}

The maximum contribution to the 
coalescence integral is obtained from the equal
mass part of the mass distributions.

The calculated meson and baryon source functions and their space-time
integrals in eq.(\ref{coal1}) yield theoretical spectra comparable
to experimental data \cite{PHENIX}. In Figure \ref{spect_pion} we display our results
for pion production together with the experimental data on
$\pi^0$ and $\pi^+$ production at mid-rapidity in Au+Au collisions
at $\sqrt{s} = 200$ AGeV. Figure \ref{spect_aprot} shows calculation
and data for antiproton. The agreement in the $0.5 < p_T < 5.0$ GeV
momentum window is astonishing and proves the validity of the coalescence 
scenario. 

Besides the hadron production on a logarithmic scale, we display the
${\overline p}^- / \pi^-$ ratio in linear scale on Figure 6.
The theoretical result is able to follow the tendencies of the
anomalous antiproton to pion ratio.

\newpage

\begin{figure}[ht]
\begin{minipage}[b]{160mm} 
\includegraphics[width=100mm,height=80mm]{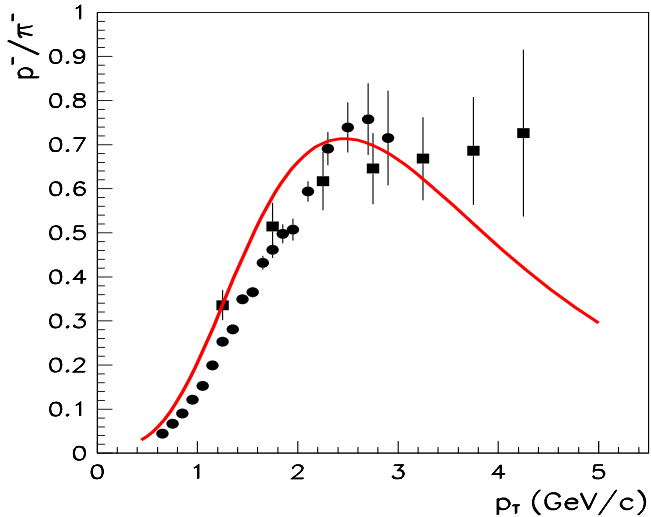}
 \caption{${\overline p}/\pi^{-}$ ratio as a function of $ p_T$
   in central Au+Au collisions at $\sqrt{s}=200$ AGeV. 
   Experimental data points are from Ref.~\cite{PHENIX}.}
\end{minipage}
\label{Fig.6}
\end{figure}

\subsection{The v2 paradox}

The meson and baryon elliptic flow at high $p_T$ 
show a peculiar behaviour, 
which is understandable in the framework of parton coalescence 
dynamics \cite{MoV}.

Assuming the following form for the asymmetric quark momentum distribution:
\begin{equation}  
  \frac{dN_q}{p_{\perp}dp_{\perp}d\Phi} =
 \frac{1}{2\pi}\frac{dN_q}{p_{\perp}dp_{\perp}}(1+2v_{2,q}(q_{\perp}) 
cos(2\Phi)) \\
\end{equation}  
one gets in the coalescence dynamics for $v_2 \ll 1$ the hadron
elliptic flow:
\begin{eqnarray}
  v_{2,M}(p_{\perp}) &\approx& 2*v_{2,q}(p_{\perp}/2) \\  
  v_{2,B}(p_{\perp}) &\approx& 3*v_{2,q}(p_{\perp}/3) \nonumber
\end{eqnarray}  

\begin{figure}[t]
\begin{minipage}[t]{145mm} 
\includegraphics[width=120mm]{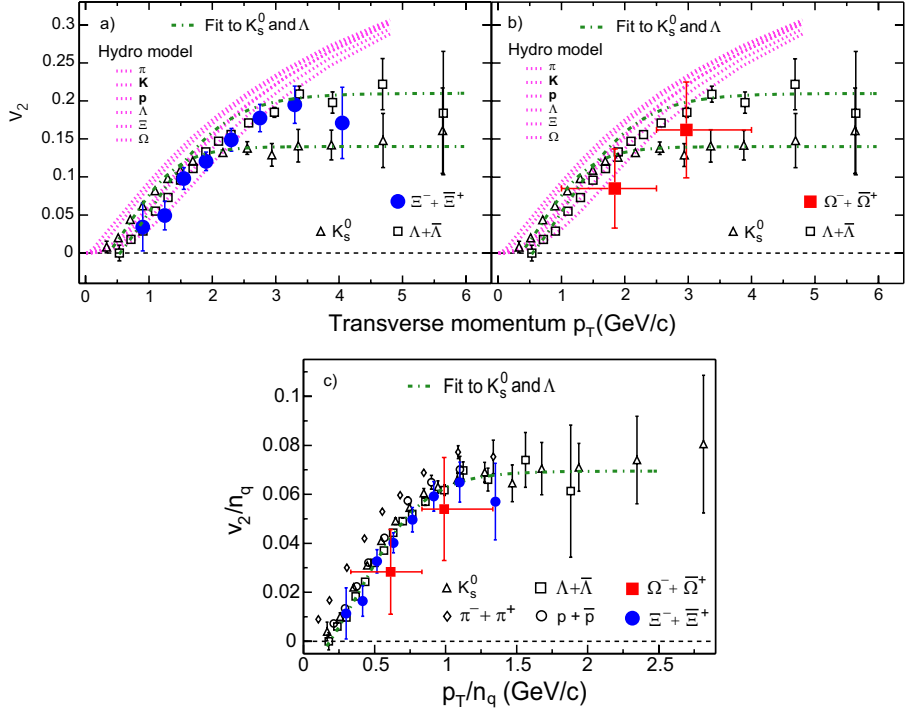}
\vspace{-1.0cm}
\caption[]{Measured baryon and meson elliptic flow 
(figure is from Ref.~\cite{v2STAR}).}
\label{Fig.7}
\end{minipage}
\end{figure}

This behaviour of the $ v_{2,M}$ and $v_{2,B} $ parameters
predicted by the coalescence hadronization dynamics 
are clearly observable in the  
experimental data (see Fig.~\ref{Fig.7}).

\noindent
The behaviour of flow asymmetry parameter
\begin{equation}
  { v_{2,u} = v_{2,d} = v_{2,s} }
\end{equation}
implies that the
 collective flow for  quarks of all flavour  
is the same. 

\newpage


\section{Conclusion }

 A large number of experimental facts is in agreement  
 with the assumption that {\it in the prehadronization stage }
\begin{itemize} 
\item the quark matter consists of 
constituent quarks and antiquarks with effective mass; 
\item the collective flow for quarks of all flavor is the same;
\item the hadronization proceeds via coalescence mechanism; 
\item the numbers of quarks and antiquarks are conserved during hadronization.
\end{itemize}
Further
\begin{itemize}
\item the initial stage of heavy ion collision is strong color field (gluon) dominated; 
\item the final stage of heavy ion collision is quark dominated. 
\end{itemize}

\end{document}